\documentclass[showpacs,twocolumn,showkeys,amsmath,amssymb,pra]{revtex4-1}

\usepackage{graphicx}   

\begin{document}

\title{Effective time-reversal via periodic shaking} 

\author{Christoph Weiss}

\affiliation{Department of Physics, Durham University, Durham DH1 3LE, United Kingdom
}
\affiliation{Institut f\"ur Physik, Carl von Ossietzky Universit\"at, D-26111 Oldenburg, Germany}

\keywords{03.75.Lm, 03.65.Xp}
                  
\date{\today}
 
\begin{abstract}
For a periodically shaken optical lattice, effective time-reversal is investigated numerically. For interacting ultra-cold atoms, the scheme of [J.\ Phys.\ B 45, 021002 (2012)] involves a quasi-instantaneous change of both the shaking-amplitude and the sign of the interaction. As the wave function returns to its initial state with high probability, time-reversal is ideal to distinguish pure quantum dynamics from the dynamics described by statistical mixtures.
\end{abstract} 
\pacs{03.75.Lm, 03.65.Xp}
\keywords{periodic driving, time-reversal, optical lattice}
\maketitle 


\section{\label{sec:introduction}Introduction}

\begin{figure}[h]
\includegraphics[width=\linewidth]{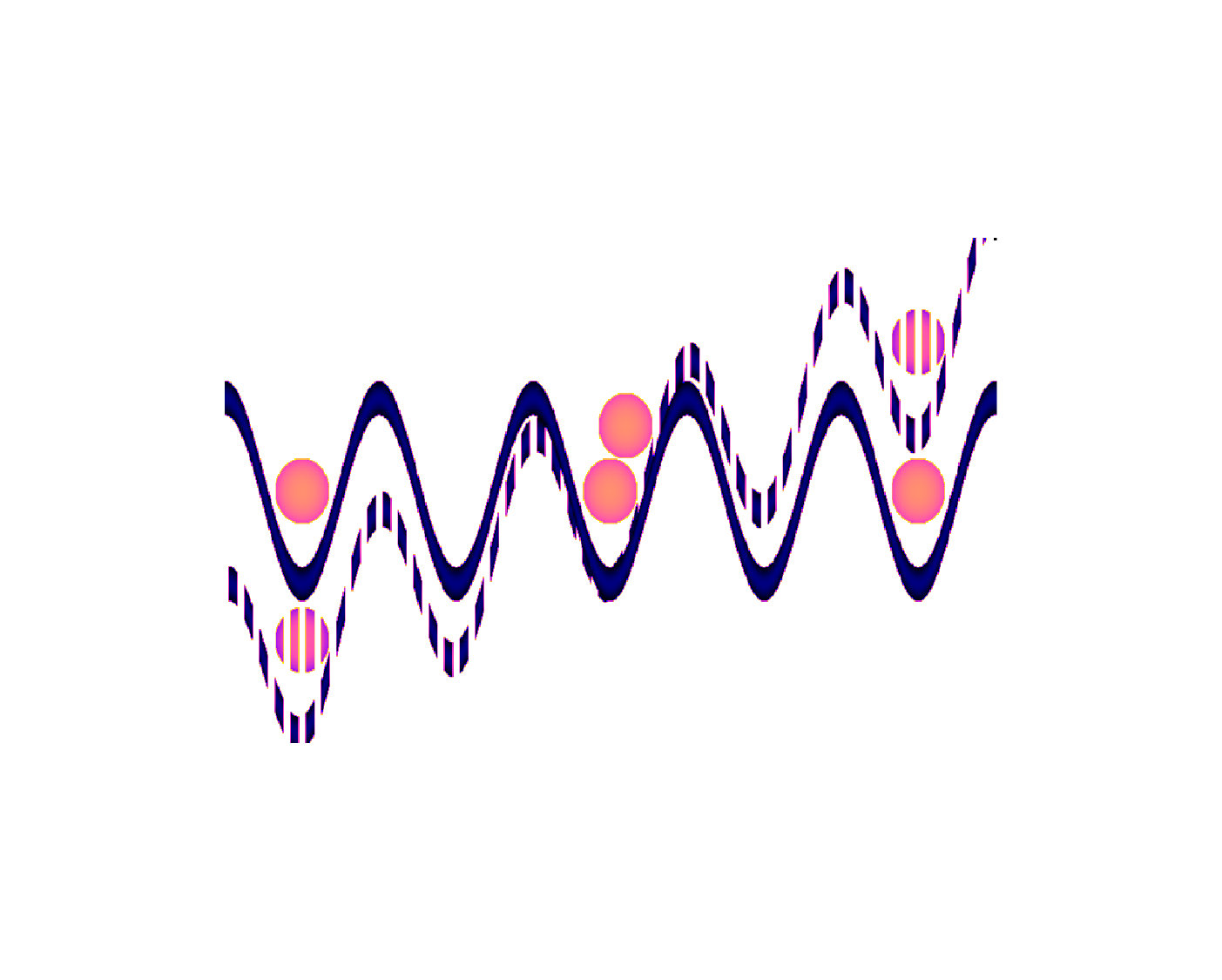}
\caption{\label{fig:sketch}(Colour online) Ultra-cold atoms in an optical lattice combined with periodic shaking. The full quantum dynamics are modelled by the time-dependent Bose-Hubbard Hamiltonian~(\ref{eq:H}); many aspects can be understood by the effective Hamiltonian~(\ref{eq:Heff}).}
\end{figure}

In the classical world our intuition is based on, there are many processes which cannot easily be reversed. For many cases a film run backwards will be easily identifiable as such: a broken glass does not reassemble itself and jump on the table; a mixture of milk and tea does not unmix in such a way that the milk flows back into the milk bottle. However, the same is not always true for quantum mechanics. This paper investigates effective time-reversal~\cite{Weiss12} via periodic shaking (Fig.~\ref{fig:sketch}) for ultra-cold bosonic atoms in an optical lattice~\cite{LewensteinEtAl07,BlochEtAl08,Yukalov09}.

Tunnelling control via periodically shaking an optical  lattice has already been applied in several experimental setups. The topics investigated experimentally range from control of the superfluid-to-insulator transition~\cite{ZenesiniEtAl09} (cf.~\cite{EckardtEtAl05b,CreffieldMonteiro06}) over frustrated classical magnetism~\cite{Struck2011}  to photon-assisted tunnelling~\cite{SiasEtAl08,ChenEtAl11,MaEtAl11} (cf.~\cite{CreffieldEtAl10,TeichmannEtAl09,XieEtAl10,EsmannEtAl12}). Research on periodically shaken systems also includes destruction of tunnelling~\cite{GrossmannEtAl91,DellaValleEtAl07,KierigEtAl08,GongEtAl09}, generation of quantum superpositions~\cite{Creffield2007,StieblerEtAl11}, robust dynamical recurrences~\cite{AyubSaif12} and reflection-less defects~\cite{LonghiDellaValle11}.

Experiments with periodically shaken optical lattices~\cite{ZenesiniEtAl09,HallerEtAl10,ChenEtAl11,MaEtAl11} can often be modelled via a time-periodic Bose-Hubbard Hamiltonian like
\begin{eqnarray}
\label{eq:H}
\hat{H}(t) =&-&J\sum_{\ell=-\infty}^{\infty}\left(\hat{c}_{\ell}^{\dag}\hat{c}_{\ell +1}^{\phantom\dag}+\hat{c}_{\ell+1}^{\dag}\hat{c}_{\ell}^{\phantom\dag} \right) \nonumber\\ &+& \frac U2\sum_{\ell=-\infty}^{\infty}\hat{c}_{\ell}^{\dag}\hat{c}_{\ell}^{\dag}\hat{c}_{\ell}^{\phantom\dag}\hat{c}_{\ell}^{\phantom\dag}\nonumber\\
&+&\sum_{\ell=-\infty}^{\infty}2\ell\hbar\mu\cos(\omega t +\alpha)\hat{c}_{\ell}^{\dag}\hat{c}_{\ell}^{\phantom\dag}\;.
\end{eqnarray}
The operators $\hat{c}^{{\phantom{\dag}}}_{\ell}$/$\hat{c}^{\dag}_{\ell}$ annihilate/create a boson in well~$\ell$;
$J$ describes the hopping of particles in the lattice and $2\hbar\mu$ is the driving amplitude. The on-site pair interaction is denoted by $U$. 
In the high-frequency limit ($\hbar\omega\gg U$, $\hbar\omega\gg J$), many aspects of experiments can be understood by an effective-Hamiltonian approach for which  Eq.~(\ref{eq:H}) is replaced by~\cite{ZenesiniEtAl09}
\begin{eqnarray}
\label{eq:Heff}
\hat{H}_{\rm eff} = &-&J_{\rm eff}\sum_{\ell=-\infty}^{\infty}\left(\hat{c}_{\ell}^{\dag}\hat{c}_{\ell +1}^{\phantom\dag}+\hat{c}_{\ell+1}^{\dag}\hat{c}_{\ell}^{\phantom\dag} \right) \nonumber\\ &+& \frac U2\sum_{\ell=-\infty}^{\infty}\hat{c}_{\ell}^{\dag}\hat{c}_{\ell}^{\dag}\hat{c}_{\ell}^{\phantom\dag}\hat{c}_{\ell}^{\phantom\dag}\;,
\end{eqnarray}
where the hopping $J$ is multiplied by the ${\cal J}_0$ Bessel-function~\cite{Abramowitz84}
\begin{equation}
\label{eq:Jeff}
J_{\rm eff}\equiv J{\cal J}_{0}\left({\textstyle\frac{2\mu}{\omega}}\right).
\end{equation}
For larger interactions, the model~(\ref{eq:Heff}) has to be modified (cf.~\cite{CreffieldMonteiro06,EsmannEtAl12}).

In experiments, systems modelled by time-periodic Hamiltonians can be switched on quasi-instantaneously. If  -- as in Eq.~(\ref{eq:H}) -- the Hamiltonian contains a term $\propto \cos(\omega t+\alpha)$, it does make a difference if the cosine is at a minimum or maximum. Counter-intuitively, switching it on at a maximum is the best thing to do if one does not want to change the wave function (cf.~\cite{RidingerDavidson07,RidingerWeiss09}). For periodically shaken systems of ultra-cold atoms, sudden changes of shaking have been used to cover topics as diverse as super Bloch oscillations~\cite{HallerEtAl10,KudoMonteiro11,ArlinghausHolthaus11}, directed transport~\cite{CreffieldSols11}, effective magnetic fields~\cite{Kolovsky11} and time-reversal for a Bose-Einstein condensate in a double well~\cite{Weiss12}.

The topic of this paper is to apply the time-reversal scheme of Ref.~\cite{Weiss12} to periodically shaken lattices. Section~\ref{sec:quantum} motivates time-reversal as a method to distinguish quantum dynamics from similar looking classical diffusion. The model used to describe the shaken lattice is introduced in Sec.~\ref{sec:model} (and in the Appendix). Section~\ref{sec:sudden} introduces the time-reversal scheme. The results are presented in Sec.~\ref{sec:results}. 

\section{\label{sec:quantum}Quantum dynamics vs.\ classical diffusion}
\begin{figure}
\includegraphics[width=\linewidth]{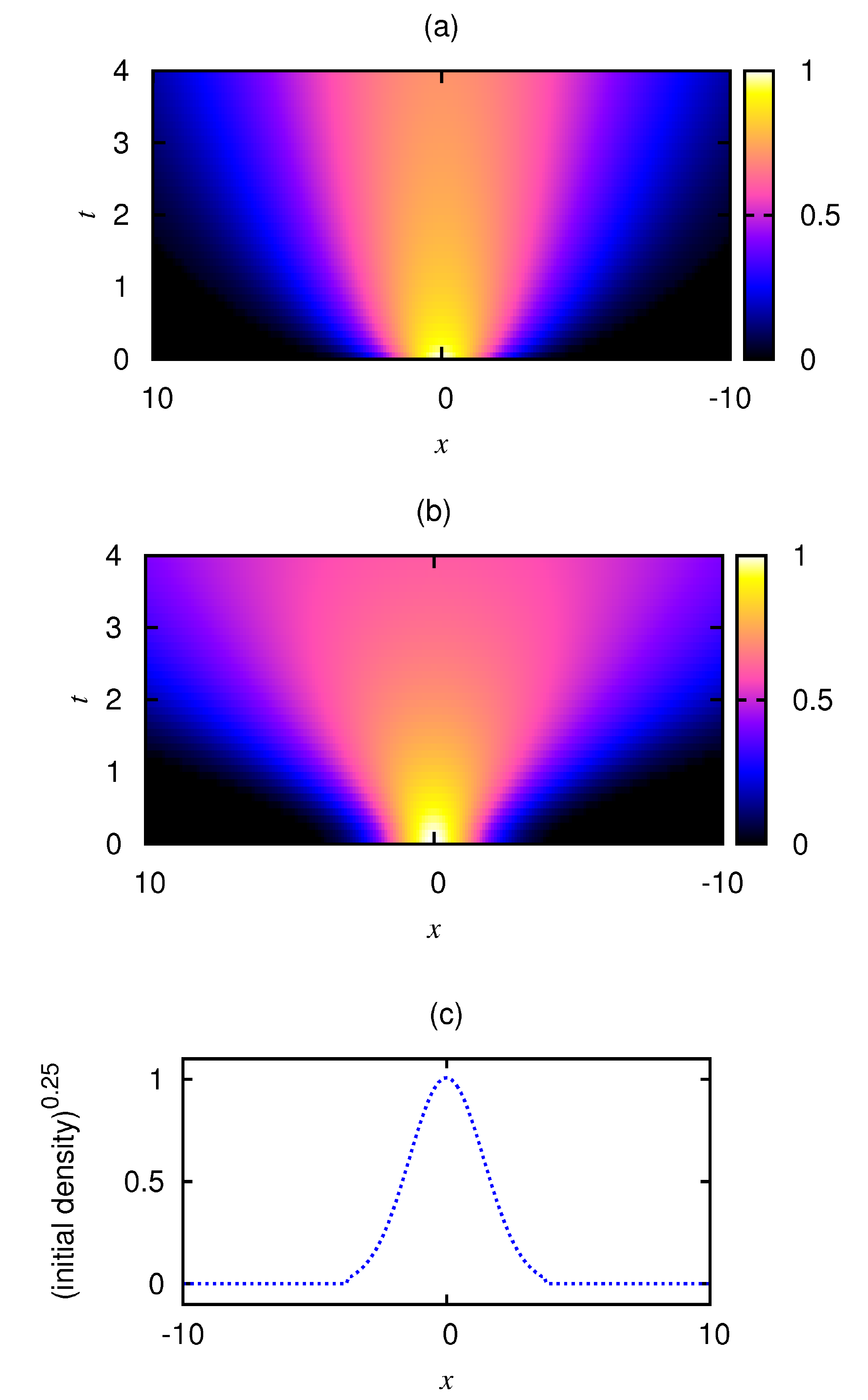}
\caption{\label{fig:diffquant}{Quantum dynamics versus classical dynamics in one dimension.} \textbf{(a)} Two-dimensional projection of the solution $\varrho(x,t)$ of the classical diffusion equation~(\ref{eq:diff}). 
To increase the visibility, $\varrho^{0.25}$ was plotted (in arbitrary units). \textbf{(b)} Two-dimensional projection of the probability density  $|\Psi(x,t)|^2$ of a free particle [cf.~Eq.~(\ref{eq:sch})]. 
Again, $(|\Psi|^2)^{0.25}$ is plotted. \textbf{(c)}~The initial conditions for both cases are chosen to be the same.}
\end{figure}

Figure~\ref{fig:diffquant} shows both the solution of the one-dimensional diffusion equation and of the one-dimensional Schr\"odinger equation for a free particle. 
For Fig.~\ref{fig:diffquant} both the classical diffusion equation in one dimension~\cite{MathewsWalker1970},
\begin{equation}
\label{eq:diff}
\partial_t
\varrho(x,t) = D
\partial_x^2
\varrho(x,t)\,,
\end{equation}
where $D>0$ is the diffusion constant,
and the Schr\"odinger equation for a free particle of mass $m$ in one dimension,
\begin{equation}
\label{eq:sch}
i\hbar
\partial_t
\Psi(x,t) = 
-\frac{\hbar^2}{2m}
\partial_x^2
\Psi(x,t)\,,
\end{equation}
have been solved for the initial conditions $\varrho(x,t\!=\!0) = \exp(-{x}^{2})/{\sqrt{\pi }}$ and 
$\Psi(x,t\!=\!0) = \exp(-0.5\,{x}^{2})/{\sqrt [4]{\pi }}$ using dimensionless variables ($D=1$, $m=0.5$, $\hbar=1$). Both the solution of the diffusion equation~\cite{MathewsWalker1970} and the solution of the Schr\"odinger equation~\cite{Fluegge90} are known analytically.
The rate at which the width of these distributions spreads is a valuable way to distinguish quantum dynamics from classical dynamics~\cite{SteinigewegEtAl07}.

On the quantum level it is, in principle, possible to construct an initial
wave function for which the width initially decreases by, e.g., numerically preparing a wave function for which the probability function reverses the motion shown in Fig.~\ref{fig:diffquant}~(b). 
There is even an example where such a case was realised experimentally: by cleverly changing the wave function~\cite{MorigiEtAl02}, the quantum motion has been turned backwards in time for atom-field interaction in a cavity quantum electrodynamics experiment~\cite{Meunier05}. Time-reversal schemes on the level of the Gross-Pitaevskii equation can be found in Refs.~\cite{MartinEtAl08,TsukadaEtAl08}.

\section{\label{sec:model}Model}

This section derives the effective Hamiltonian~(\ref{eq:Heff}) for a single particle in a double well; the extension to an optical lattice is given in the Appendix.

For a single particle in a double well, the Hamiltonian~(\ref{eq:H}) reduces to
\begin{eqnarray}
\label{eq:Hs}
\hat{H}(t) &=& -J\left(\hat{c}_1^{\dag}\hat{c}_2^{\phantom\dag}+\hat{c}_2^{\dag}\hat{c}_1^{\phantom\dag} \right) 
\nonumber\\
&&+\hbar\mu\cos(\omega t +\alpha)\left(\hat{c}_2^{\dag}\hat{c}_2^{\phantom\dag}-\hat{c}_1^{\dag}\hat{c}_1^{\phantom\dag}\right)\;;
\end{eqnarray}
one thus has to solve the corresponding Schr\"odinger equation:
\begin{eqnarray}
 i \partial_t\psi_1(t) &=& -\mu \cos(\omega t+\alpha)\psi_1(t) -\frac{J}{\hbar} \psi_2(t)\\
 i \partial_t\psi_2(t) &=& +\mu \cos(\omega t+\alpha)\psi_2(t) -\frac{J}{\hbar} \psi_1(t)
\end{eqnarray}
The ansatz
\begin{eqnarray}
\label{eq:ansatz1}
 \psi_1(t) &=& \varphi_1(t)\exp\left[i\frac{\mu}{\omega}\sin(\omega t+\alpha)\right] \\\label{eq:ansatz2}
 \psi_2(t) &=& \varphi_2(t)\exp\left[-i\frac{\mu}{\omega}\sin(\omega t+\alpha)\right]
\end{eqnarray}
leads to
\begin{eqnarray}
 i \partial_t\varphi_1(t) &=& -\frac{J}{\hbar} \varphi_2(t)\exp\left[i\frac{2\mu}{\omega}\sin(\omega t+\alpha)\right]\\
 i \partial_t\varphi_2(t) &=& -\frac{J}{\hbar} \varphi_1(t)\exp\left[-i\frac{2\mu}{\omega}\sin(\omega t+\alpha)\right].
\end{eqnarray}
In order to solve this in the high-frequency limit $\hbar\omega\gg J$ (which may already be reached for $\hbar\omega\approx 10J$), the expansion in Bessel functions~\cite{Abramowitz84} is particularly useful
\begin{equation}
\label{eq:bessel}
e^{iz\sin(\omega t+\alpha)}=\sum_{k=-\infty}^{\infty}{\cal J}_{k}(z)e^{ik(\omega t+\alpha)}\,.
\end{equation}
For high frequencies, only the time-independent part of this sum is relevant in the above equations, leading to:
\begin{eqnarray}
\label{eq:eff1pt1}
 i \partial_t\varphi_1(t) &\simeq& -\frac{J_{\rm eff}}{\hbar} \varphi_2(t)\\
 i \partial_t\varphi_2(t) &\simeq& -\frac{J_{\rm eff}}{\hbar} \varphi_1(t),
\label{eq:eff1pt2}
\end{eqnarray}
where $J_{\rm eff}$ is given by Eq.~(\ref{eq:Jeff}).

The effective tunnelling can be both zero and negative. Note that sometimes the physics is described by a zero~\cite{ZenesiniEtAl09} whereas in other cases the dynamics is governed by the other Bessel functions~\cite{EsmannEtAl12}.
Equations~(\ref{eq:eff1pt1}) and (\ref{eq:eff1pt2}) yield the effective, time-independent Hamiltonian:
\begin{equation}
\label{eq:Heff2}
\hat{H}_{\rm eff} = - J_{\rm eff} \left(\hat{c}_1^{\dag}\hat{c}_2^{\phantom\dag}+\hat{c}_2^{\dag}\hat{c}_1^{\phantom\dag} \right)\;.
\end{equation} 
Repeating the above derivation for the case of an optical lattice (see Appendix) gives the effective Hamiltonian~(\ref{eq:Heff}) for non-interacting particles. It remains valid even for interacting particles, as long as the interaction is not too large.


Strictly speaking, in the periodically shaken case there is no lowest energy: energies are only defined up to integer multiples of $\hbar \omega$. Thus, the quasi-energies for high frequencies read:
\begin{equation}
 \epsilon= \pm J_{\rm eff}+j \hbar \omega\;,
\end{equation}
with integer $j$.

For negative effective tunnelling, the ground state wave function in the effective model is antisymmetric, which would be impossible without shaking. Despite the infinite number of quasi-energies, there are only two Floquet-states for the case discussed here:
\begin{eqnarray}
\psi_{\pm}(t)&\simeq&\left(\begin{array}{c}\nonumber
\varphi_1(t)\\
\varphi_2(t)
\end{array}\right)_{\pm}\\
&=&\frac{e^{\pm i J_{\rm eff}t/\hbar}}{\sqrt{2}}\left(\begin{array}{c}
\exp[i\frac{\mu}{\omega}\sin(\omega t+\alpha)]\\
\pm \exp[-i\frac{\mu}{\omega}\sin(\omega t+\alpha)]
\end{array}\right)
\label{eq:floquedouble}
\end{eqnarray}

In order to calculate the time-averaged energy in the time-dependent Hamiltonian,
\begin{equation}
\label{eq:hmitteldef}
\overline{\langle \hat{H}(t)\rangle}_{\pm}\equiv\frac1T\int_0^T\langle\psi_{\pm}(t)|\hat{H}|\psi_{\pm}(t)\rangle\;,
\end{equation}
one needs
\begin{eqnarray}
\langle\psi_{\pm}(t)|\hat{H}|\psi_{\pm}(t)\rangle &=& -\hbar\mu \cos(\omega t+\alpha)\psi_1^*(t)\psi_1(t)\nonumber\\\nonumber&& -{J} \psi_1^*(t)\psi_2(t)\\\nonumber
  &&+ \hbar\mu \cos(\omega t+\alpha)\psi_2^*(t)\psi_2(t)\\&& -{J} \psi_2^*(t)\psi_1(t)\;.
\end{eqnarray}
The second step involves the time-average over one oscillation period, which removes the parts proportional to  $\psi_j^*(t)\psi_j(t)$, $j=1,2$ and leaves:
\begin{eqnarray}
\overline{\langle \hat{H}(t)\rangle}_{\pm}  &=& -\frac1T\int_0^T{J} \psi_2^*(t)\psi_1(t)\nonumber\\
&&-\frac1T\int_0^T{J} \psi_1^*(t)\psi_2(t)\nonumber\\
&=&\mp J_{\rm eff},
\end{eqnarray}
which uses Eq.~(\ref{eq:bessel}).
Note that the only approximation in the above equation involves the high-frequency approximation for the Floquet-states $\psi_{\pm}(t)$.


\section{\label{sec:sudden}Sudden changes of the Hamiltonian}
\subsection{\label{sub:amplitude}Forcing time-reversal via changes to the shaking amplitude}

If the Hamiltonian is changed quasi-instantaneously, this may involve changing, for example, either the amplitude or the phase or both at $t=t_0$:
\begin{eqnarray}
\mu\to \widetilde{\mu}_1\\
\alpha\to \widetilde{\alpha}.
\end{eqnarray}
To understand the effect of sudden changes of the Hamiltonian, the example of a single particle in a double well is again a good starting point. 
In this case, the eigenfunctions will be given by
\begin{widetext}
\begin{eqnarray}
\psi_{\pm}(t)
&=&\left\{
\begin{array}{ccc}
\frac{e^{\pm i J_{\rm eff}t/\hbar}}{\sqrt{2}}\left(\begin{array}{c}
\exp\left[i\frac{\mu}{\omega}\sin(\omega t+\alpha)\right]\\
\pm \exp\left[-i\frac{\mu}{\omega}\sin(\omega t+\alpha)\right]
\end{array}\right)&:&t\le t_0\\
\phantom{a}&\phantom{a}&\phantom{a}\\
\frac{e^{\pm i J_{\rm eff}t/\hbar}}{\sqrt{2}}\left(\begin{array}{c}
\exp\left[i\frac{\widetilde{\mu}_1}{\omega}\sin(\omega t+\widetilde{\alpha})\right]\\
\pm \exp\left[-i\frac{\widetilde{\mu}_1}{\omega}\sin(\omega t+\widetilde{\alpha})\right]
\end{array}\right)&:&t> t_0
\end{array}\right.
\end{eqnarray}
\end{widetext}
Choosing the example $\frac{\mu}{\omega}=\frac{\pi}2$  shows that a particle which is initially in the $\psi_+$-state might be in the $\psi_-$-state at $t>t_0$ if the jump takes place for $\sin(\omega t_0+\alpha)=1$ and if $\widetilde{\alpha}$ is chosen such that
$\sin(\omega t_0+\widetilde{\alpha})=0$. Thus, although the wave function is continuous at $t=t_0$, the average energy can change even if the effective Hamiltonian is the same before and after the jump (cf.~\cite{RidingerWeiss09,ClearyEtAl10}).

In the following, a quasi-instantaneous change of the Hamiltonian is used to switch the sign of the time-independent Hamiltonian (\ref{eq:Heff}). The sign can be switched by quasi-instantaneously changing both the tunnelling term by switching the shaking amplitude, e.g.~\cite{Weiss12},
\begin{eqnarray}
{\mathcal J}_0(1.692)&\simeq& 0.403 \quad {\rm and}\\
{\mathcal J}_0(3.832)&\simeq& -0.403\;,
\end{eqnarray}
and the sign of the interaction via a Feshbach-resonance~\cite{BauerEtAl09}.
\begin{eqnarray}
\label{eq:hdream}
\hat{H}_{\rm ideal} &\equiv& \left\{  
\begin{array}{lcr}
  +\hat{H}_{\rm eff}&:&0\le \tau < \tau_0\\
  -\hat{H}_{\rm eff}&:&\tau\ge \tau_0
  \end{array}\;,\right.\; \tau\equiv \frac{tJ}{\hbar}\;.
\end{eqnarray} 
The corresponding unitary time-evolution is given by $U(0,\tau) = \exp[i(\tau-2\tau_0) \hat{H}_{\rm eff}/J]$ for $\tau>\tau_0$ with perfect return to the initial state at $\tau=2\tau_0$. The turning point $\tau_0$ however has to be chosen with care: only by taking $\tau_0$ close to the maximum of the shaking, can unwanted excitations be excluded (cf., e.g., Ref.~\cite{RidingerWeiss09}). In the following, the time-reversal is demonstrated by numerically solving the full, time-dependent Hamiltonian~(\ref{eq:H}) using the Shampine-Gordon routine~\cite{ShampineGordon75}.

\subsection{\label{sub:complex}Complex $J_{\rm eff}$ induced via phase-jumps}
As the ansatz in Eqs.~(\ref{eq:ansatz1}) and (\ref{eq:ansatz2}) is not unique, one can also derive effective equations for which $J_{\rm eff}$ becomes complex~\cite{CreffieldSols11,Kolovsky11,EsmannEtAl12}. Choosing, e.g., complex phase-factors such that the exponential in the ansatz is 1, i.e., 
\begin{eqnarray}
\nonumber
 \psi_1(t) &=& \varphi_1(t)\exp\left[i\frac{\mu}{\omega}\sin(\omega t+\alpha)-i\frac{\mu}{\omega}\sin(\alpha)\right] \\\nonumber
 \psi_2(t) &=& \varphi_2(t)\exp\left[-i\frac{\mu}{\omega}\sin(\omega t+\alpha)+i\frac{\mu}{\omega}\sin(\alpha)\right]\;,
\end{eqnarray}
would replace the effective Hamiltonian (\ref{eq:Heff2}) by
\begin{eqnarray}
\label{eq:Geff}
\hat{G}_{\rm eff} &=& - J_{\rm eff} \left(e^{i\frac{\mu}{\omega}\sin(\alpha)}\hat{c}_1^{\dag}\hat{c}_2^{\phantom\dag}+e^{-i\frac{\mu}{\omega}\sin(\alpha)}\hat{c}_2^{\dag}\hat{c}_1^{\phantom\dag} \right)
\end{eqnarray}
where $J_{\rm eff}$ still is the real number defined in Eq.~(\ref{eq:Jeff}).

The corresponding $2\times 2$-matrix has the same eigenvalues but different eigenvectors than Eq.~(\ref{eq:Heff2}). However, the full time-dependent solution is again identical to Eq.~(\ref{eq:floquedouble}). Thus, as before, the ground state does not depend on the phase of the driving. Choosing 
\begin{equation}
\label{eq:jump}
e^{i\frac{\mu}{\omega}\sin(\alpha)}=-1
\end{equation}
 rather than 1 would, however, give the impression that the sign of the effective Hamiltonian has changed and ground and excited states had swapped their places (this was the aim of Sec.~\ref{sub:amplitude}). Using the full Hamiltonian shows that is not the case. In the following, the jump is modelled at $t=0$ under the assumption that before the jump one has $\alpha =0$ and $J_{\rm eff} >0$. Introducing the phase-jump~(\ref{eq:jump}) quasi-instantaneously changes the Floquet-states 
\begin{eqnarray}
|\psi_+(t)\rangle_{\alpha = 0} &\longrightarrow& |\psi_-(t)\rangle_{\alpha \ne 0}\\
|\psi_-(t)\rangle_{\alpha = 0} &\longrightarrow& |\psi_+(t)\rangle_{\alpha \ne 0}\;,
\end{eqnarray}
where $\alpha\ne 0$ refers to the value defined by Eq.~(\ref{eq:jump})
but leaves the time-averaged energies~(\ref{eq:hmitteldef}) unchanged:
\begin{eqnarray}
\overline{\langle \hat{H}(t)\rangle}_+&=& {\rm const.}\\
\overline{\langle \hat{H}(t)\rangle}_-&=& {\rm const.}
\end{eqnarray}
Contrary to this, the sudden change of the amplitude discussed in Sec.~\ref{sub:amplitude} leaves the \textit{wave functions unchanged} but \textit{quasi-instantaneously changes the sign of the average energy}. This leads, in particular, to having a ground state which is antisymmetric. Another example of a related system with a counter-intuitive ground state is discussed in \cite{EckardtEtAl10}. As the ground-state properties do not change by the jump~(\ref{eq:jump}), this paper avoids calling Eq.~(\ref{eq:Geff}) an effective Hamiltonian. 

As long as one is primarily interested in the dynamics (rather than, e.g., the ground state) it would nevertheless be perfectly valid to use such equations; the decision if phase-jumps or jumps in the shaking amplitude are preferable is a question of available technical laboratory resources and thus can not be discussed by the present paper. While this paper focuses on sudden changes for which $\sin(\alpha)=0$, discussing $\sin(\alpha)\ne 0$ can be easier for Eq.~(\ref{eq:Geff}) rather than Eq.~(\ref{eq:Heff2}) as the latter will involve complicated changes in the effective wave function at the time of the jump.

\begin{figure}
\includegraphics[width=\linewidth]{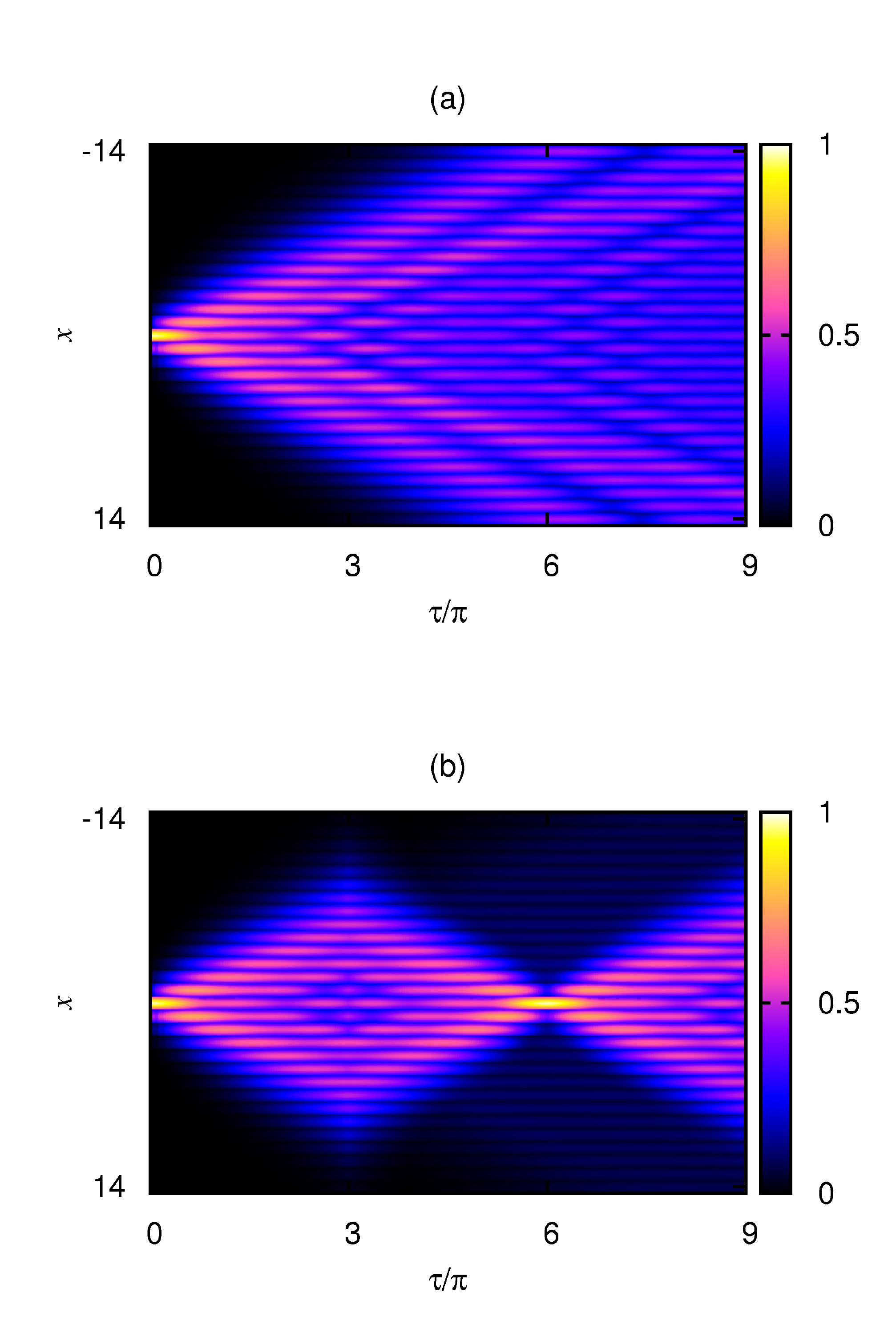}
\caption{\label{fig:gitter}(Colour online) {Time-reversal of the quantum dynamics for two interacting particles in a periodically shaken optical lattice.} \textbf{(a)} Two-dimensional projection of the probability density (to the power $1/4$, in arbitrary units) if both particles initially are close to the middle of the lattice ($U=1.6J$, $\hbar\omega = 20J$, $2\mu/\omega=1.692\,J$).  \textbf{(b)} After applying coherently controlled time-reversal at $\tau=\tau_0$, the ``echo'' is visible near $\tau\approx 6\pi$ ($\tau_0 = 3\pi$; all parameters as in a except for $\tau\ge \tau_0$: $2\mu/\omega=3.832\,J$ and $NU = -1.6J$).}
\end{figure}
\section{\label{sec:results}Results}

Figure~\ref{fig:gitter} shows the spreading of the probability density for two interacting particles initially localised near the centre of a periodically shaken optical lattice. 
The initial state was chosen to be the ground state of a three-lattice-site version of the effective Hamiltonian~(\ref{eq:Heff}) where the potential energy of the two outer wells was increased by $10J$ compared to the middle well, thus mimicking the strong harmonic confinement of a Bose-Einstein condensate released into such a shaken lattice in Ref.~\cite{LignierEtAl07}. To visualise the probability density (which is a single number for each lattice site), a Gaussian density profile of the Wannier-function was modelled.

\begin{figure}[!ht] 
\includegraphics[width=\linewidth]{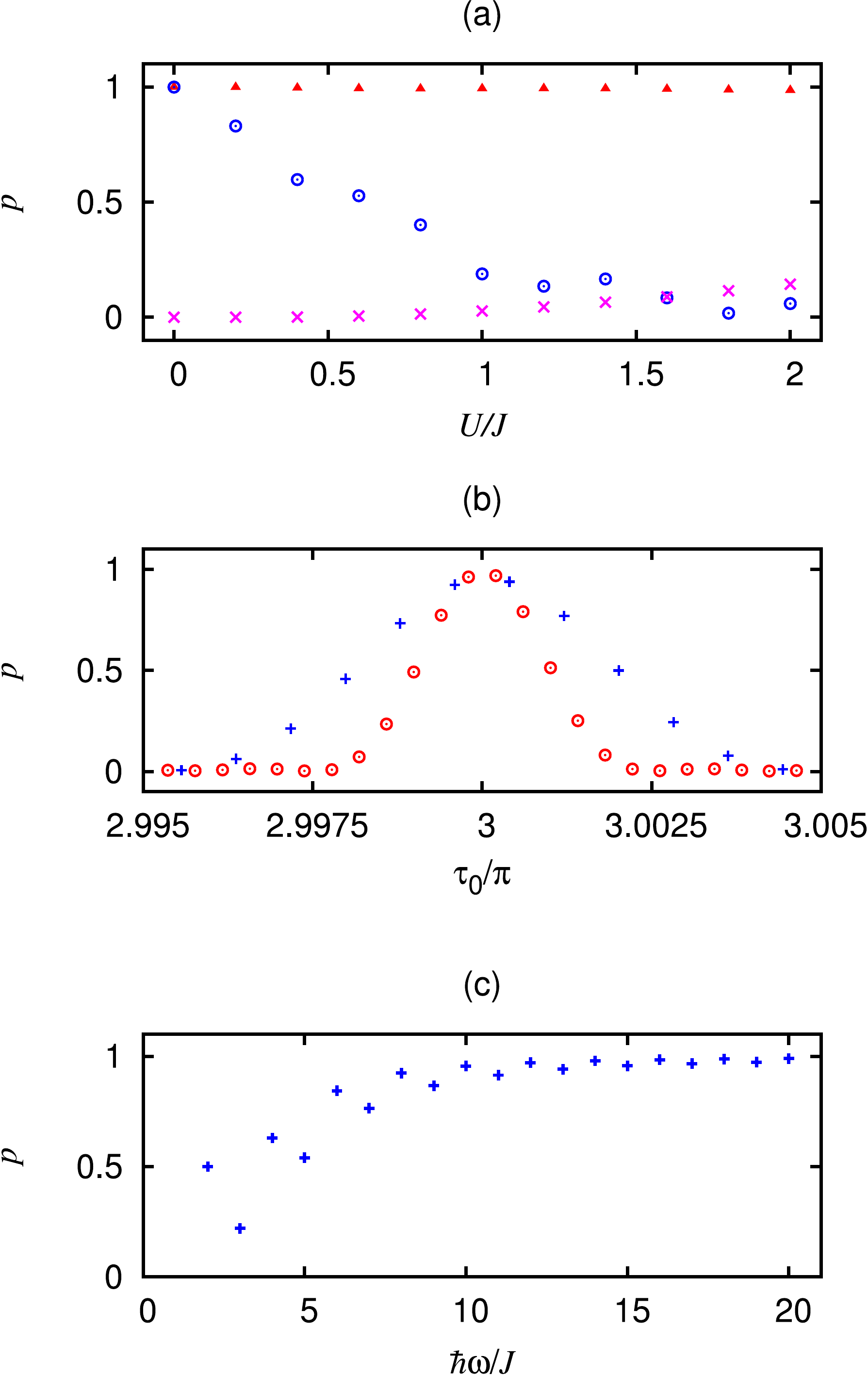}
\caption{\label{fig:gitterscan}(Colour online) {Time-reversal in a periodically shaken optical lattice} \textbf{(a)} The return probability~(\ref{eq:return}) at $\tau=2\tau_0$ as a function of interaction. It is only large if both interaction and amplitude are changed (red/triangles). Switching 
only the shaking amplitude (blue/circles) as expected only works for non-interacting particles, switching only the interaction (magenta/crosses) may work for high interactions. \textbf{(b)} The return probability as a function of switching time has a broader distribution for low frequencies  (blue/crosses, $\hbar \omega=10 J$)  than for high frequencies (red/circles, $\hbar\omega =20 J$). All other parameters are as in panel (b). \textbf{(c)} The return probability as a function of shaking frequency decreases for lower angular frequencies (all other parameters as in b). While in the high frequency regime [cf.\ Eqs.~(\ref{eq:floquedouble}) and (\ref{eq:floquelattice})] instantaneously changing the Hamiltonian should be equally reliable for any zero of $\sin(\omega t+\alpha)$; for lower frequencies multiples of $2\pi$ lead to better return probabilities than $\pi$, $3\pi$, $\ldots$ (cf.~\cite{RidingerWeiss09}).}
\end{figure}

The dynamics displayed in Fig.~\ref{fig:gitter}~(a) are similar to the text-book case of a free particle in one-dimension~\cite{Fluegge90}. At first glance, the main difference to the classical diffusion equation seems to be the rate at which the probability density expands (cf.~\cite{SteinigewegEtAl07}). The time-reversal visible in Fig.~\ref{fig:gitter}~(b) demonstrates deeper differences between classical and quantum physics. To quantify the quality of time-reversal, the ``return probability'', i.e., the probability to return to the original state,
\begin{equation}
\label{eq:return}
p\equiv\left|\langle \Psi(\tau\!=\!0)|\Psi(\tau\!=\!2\tau_0)\rangle\right|^2\;,
\end{equation}
is used (Fig.~\ref{fig:gitterscan}).  This figure shows that, as for the case of the Bose-Einstein condensate in a periodically shaken double well~\cite{Weiss12}, the time-reversal is not too sensitive to the precise choice of experimentally relevant parameters like interaction [Fig.~\ref{fig:gitterscan}~(a)], point of switching [Fig.~\ref{fig:gitterscan}~(b)] and shaking frequency [Fig.~\ref{fig:gitterscan}~(c)].

The high probability to return to the initial state observed in the numerics offers an ideal method to distinguish between statistical mixtures and pure quantum states (cf.~\cite{Weiss12}). For larger particle numbers, 
experiments could observe the ``echo'' by studying the width of the wave function with the existing experimental setups of Refs~\cite{HallerEtAl10,LignierEtAl07}.

\section{\label{sec:concl}Conclusion}

To conclude, effective time-reversal was numerically  induced via periodic shaking to an optical lattice for two indistinguishable interacting bosons. The time-reversal was ideally described by an effective Hamiltonian and subsequently tested numerically by using the full, time-dependent model. 

Time-reversal was realised by switching the sign of the effective Hamiltonian by quasi-instantaneously changing the strength of the shaking. At the same time, the scheme suggests a change of the sign of the interaction via a Feshbach resonance. The Floquet-states remain unchanged by this; however their time-averaged mean energy switches its sign. This reflects the change of the sign of the effective Hamiltonian and supports the claim that the effective equations used to describe the dynamics can be labelled as an effective Hamiltonian. 

The high probability to return to the initial state shows that the time-reversal can be used to distinguish quantum mechanics from statistical mixtures induced by decoherence.

The numerical simulations indicate that the effective time-reversal should be realisable with present day technology.
To further optimise suitable time-reversal in an experiment, the feed-back loops of optimal  control theory could be used~\cite{JudsonRabitz92,AssionEtAl98}.

\acknowledgments
I thank S.\ A.\ Gardiner, E.\ Haller, M.\ Holthaus, S.\ Trotzky and C.\ Vaillant for discussions.

\begin{appendix}
\section{\label{sub:single}Effective Hamiltonian for a single particle in a tight-binding lattice}
The Schr\"odinger equation for a single particle in a shaken lattice reads:
\begin{eqnarray}
 i \partial_{\ell}\psi_{\ell}(t) &=& 2\ell\mu \cos(\omega t+\alpha)\psi_{\ell}(t) \\\nonumber
&-&\frac{J}{\hbar} \left[\psi_{\ell+1}(t)+\psi_{\ell-1}(t)\right]\;,
\end{eqnarray}
with integer $\ell$.  
Choosing the ansatz
\begin{eqnarray}
\label{eq:ansatzg1}
 \psi_{\ell}(t) &=& \varphi_{\ell}(t)[A(t)]^{\ell}\;,
\end{eqnarray}
where 
\begin{equation}
A(t)=\exp\left[-i\frac{2\mu}{\omega}\sin(\omega t+\alpha)\right]\;,
\end{equation}
leads to
\begin{eqnarray}
 i \partial_{\ell}\varphi_{\ell}(t)=-\frac{J}{\hbar} \left[A(t)\varphi_{\ell+1}(t)+[A(t)]^{-1}\varphi_{\ell-1}(t)\right].
\end{eqnarray}
Within the high-frequency regime, one gets:
\begin{eqnarray}
 i \partial_{\ell}\varphi_{\ell}(t)\simeq-\frac{J_{{\rm eff}}}{\hbar} \left[\varphi_{\ell+1}(t)+\varphi_{\ell-1}(t)\right],
\end{eqnarray}
where $J_{{\rm eff}}$ is again given by Eq.~(\ref{eq:Jeff}).

Thus, the effective Hamiltonian now reads:
\begin{equation}
\label{eq:HeffgU0}
\hat{H}_{\rm eff} = -J_{\rm eff}\sum_{\ell=-\infty}^{\infty}\left(\hat{c}_{\ell}^{\dag}\hat{c}_{\ell +1}^{\phantom\dag}+\hat{c}_{\ell+1}^{\dag}\hat{c}_{\ell}^{\phantom\dag} \right) 
\end{equation}

The eigenfunctions of this effective Hamiltonian are known:
\begin{equation}
|\varphi_{k}(t)\rangle=\exp\left(-i\frac{\varepsilon(k)}{\hbar}t\right)\sum_{\ell=-\infty}^{\infty}\exp(ikd\ell)|\ell\rangle\;,
\end{equation}
where $d$ is the lattice constant, $k$ the quasi-momentum and the energy is
\begin{equation}
\varepsilon(k) = -2J_{\rm eff} \cos(kd)\;.
\end{equation}
The notation $\varepsilon(k)$  was chosen to indicate that it will become the quasi-energy of the time-dependent problem.

 Within the high-frequency approximation, the Floquet-states are now:
\begin{equation}
\label{eq:floquelattice}
|\psi_{k}(t)\rangle=\exp\left(-i\frac{\varepsilon(k)}{\hbar}t\right)\sum_{\ell=-\infty}^{\infty}[A(t)]^{\ell}\exp(ikd\ell)|\ell\rangle.
\end{equation}
In order to calculate the mean energy, time-averaged over one oscillation period $T$, where $\omega T=2\pi$  (i.e., the energy associated with this Floquet-state) we can replace the sum $\sum_{\ell=-\infty}^{\infty}$ by $\frac1{\sqrt{N}}\sum_{\ell=1}^{N}$ and take the limit $N\to\infty$ at the end. Thus, with this normalised version of $|\psi_{k}(t)\rangle$ defined on $N$ lattice sites we can write:
\begin{eqnarray}
\overline{E_k} &=& \frac1T\int_0^T{\rm d}t\,\langle\psi_{k}(t)|\hat{H}(t)|\psi_{k}(t)\rangle\\
&=&
 \lim_{N\to\infty}\frac1N\sum_{\ell=1}^N\left\{\frac1T\int_0^T{\rm d}t\,2\ell\hbar\mu\cos(\omega t +\alpha)\right.\nonumber\\
&& \left. -\frac1T\int_0^T{\rm d}t\,J\left(A(t)\exp(ikd)+[A(t)\exp(ikd)]^{-1}\right)\right\}\nonumber.
\end{eqnarray}
Without any further approximation this leads to:
\begin{eqnarray}
\overline{E_k} &=&-2J{\cal J}_0\left({\textstyle \frac{2\mu}{\omega}}\right)\cos(kd)\\
             &=&-2J_{\rm eff}\cos(kd)\nonumber\\
 &=& \varepsilon(k)\;.
\end{eqnarray}

\end{appendix}
%

\end{document}